\documentclass[sigconf]{acmart}

\usepackage{enumitem}
\usepackage{graphicx}
\usepackage{balance}
\usepackage{subcaption}
\usepackage{amsmath}
\usepackage{lineno}
\usepackage{color}
\usepackage{xcolor}
\usepackage{booktabs}
\usepackage{arydshln}
\usepackage{chngcntr}

\usepackage{multirow}

\AtBeginDocument{%
  \providecommand\BibTeX{{%
    \normalfont B\kern-0.5em{\scshape i\kern-0.25em b}\kern-0.8em\TeX}}}

\copyrightyear{2025}
\acmYear{2025}
\setcopyright{rightsretained}
\acmConference[CHIGreece 2025]{CHIGreece 2025: 3rd International Conference of the ACM Greek SIGCHI Chapter}{September 24--26, 2025}{Hermoupolis, Syros, Greece}
\acmBooktitle{CHIGreece 2025: 3rd International Conference of the ACM Greek SIGCHI Chapter (CHIGreece 2025), September 24--26, 2025, Hermoupolis, Syros, Greece}
\acmPrice{}
\acmDOI{10.1145/3749012.3749048}
\acmISBN{979-8-4007-1561-7/25/09}

\begin{document}

\title{The Reel Deal: Designing and Evaluating LLM-Generated Short-Form Educational Videos}

\author{Lazaros Stavrinou}
\orcid{0009-0004-1178-9681}
\affiliation{
  \institution{University of Cyprus}
  \city{Nicosia}
  \country{Cyprus}}
\email{stavrinou.lazaros@ucy.ac.cy}

\author{Argyris Constantinides}
\orcid{0000-0003-2737-6318}
\affiliation{
  \institution{Cognitive UX LTD \& University of Cyprus}
  \city{Nicosia}
  \country{Cyprus}}
\email{argyris@cognitiveux.com}

\author{Marios Belk}
\orcid{0000-0001-6200-0178}
\affiliation{
  \institution{Cognitive UX LTD \& University of Cyprus}
  \city{Nicosia}
  \country{Cyprus}}
\email{belk@cognitiveux.com}

\author{Vasos Vassiliou}
\orcid{0000-0001-8647-0860}
\affiliation{
  \institution{University of Cyprus \& CYENS Centre of Excellence}
  \city{Nicosia}
  \country{Cyprus}}
\email{vassiliou.vasos@ucy.ac.cy}

\author{Fotis Liarokapis}
\orcid{0000-0003-3617-2261}
\affiliation{
  \institution{CYENS Centre of Excellence}
  \city{Nicosia}
  \country{Cyprus}}
\email{f.liarokapis@cyens.org.cy}

\author{Marios Constantinides}
\orcid{0000-0003-1454-0641}
\affiliation{
  \institution{CYENS Centre of Excellence}
  \city{Nicosia}
  \country{Cyprus}}
\email{marios.constantinides@cyens.org.cy}

\renewcommand{\shortauthors}{Stavrinou et al.}

\begin{abstract}
Short-form videos are gaining popularity in education due to their concise and accessible format that enables microlearning. Yet, most of these videos are manually created. Even for those automatically generated using artificial intelligence (AI), it is not well understood whether or how they affect learning outcomes, user experience, and trust. To address this gap, we developed ReelsEd, which is a web-based system that uses large language models (LLMs) to automatically generate structured short-form video (i.e., reels) from lecture long-form videos while preserving instructor-authored material. In a between-subject user study with 62 university students, we evaluated ReelsEd and demonstrated that it outperformed traditional long-form videos in engagement, quiz performance, and task efficiency without increasing cognitive load. Learners expressed high trust in our system and valued its clarity, usefulness, and ease of navigation. Our findings point to new design opportunities for integrating generative AI into educational tools that prioritize usability, learner agency, and pedagogical alignment.
\end{abstract}


\maketitle

\section{Introduction}
\label{sec:introduction}
The rise of short-form video platforms such as TikTok, Instagram Reels, and YouTube Shorts has (re)shaped digital content consumption, especially among Generation Z, who tend to favor quick, visually engaging, and easily digestible formats~\cite{doloi2024reels}. This shift aligns with the pedagogical approach of microlearning, which aims to deliver short and focused learning content to meet learners' immediate needs~\cite{denojean2024microlearning, sankaranarayanan2023microlearning, taylor2022effects}. At the same time, recent advances in generative Artificial Intelligence (AI), particularly Large Language Models (LLMs), have enabled the automated creation of educational content that unlocked new opportunities for personalized learning experiences~\cite{mittal2024review, yu2023generative, pataranutaporn2021ai}.

Recent works have demonstrated the pedagogical potential of short-form video and AI-generated educational content. Studies have shown that microlearning formats can increase learner engagement, satisfaction, and knowledge retention~\cite{conde2024microlearning, kohnke2024microlearning}, while AI tools such as ChatGPT are currently used to support personalized tutoring, content generation, and learner assistance~\cite{bettayeb2024exploring, jensen2024generative}. However, the intersection of generative AI and short-form videos for educational purposes remains relatively underexplored in both Human-Computer Interaction (HCI) and education research.

Despite the increasing use of short-form videos for educational purposes and the rise of generative AI in content creation, few studies have examined how LLM-generated videos are perceived by learners when compared to human-created alternatives~\cite{netland2025comparing}. Moreover, little is known about how such content influences user experience, trust in AI, or the effectiveness of learning when consumed in highly compressed formats typical of social media. While learners may appreciate the accessibility and novelty of AI-generated videos, questions remain about the fidelity of such summaries, their ability to convey nuance, and the risk of oversimplification or factual distortion~\cite{netland2025comparing, jensen2024generative}. Furthermore, learners’ attitudes toward AI-generated educational content are not solely shaped by its utility, but also by perceived trustworthiness~\cite{peng2024human, toreini2020relationship, tahaei2023human, constantinides2024good}. This highlights the need to investigate how learners engage with and evaluate LLM-generated short-form educational content. Guided by this need, this paper makes two main contributions:

\begin{enumerate}
    \item We developed \emph{ReelsEd}, a web-based system that automatically generates short-form educational videos (``reels'') from long-form lecture content using GPT-4 while preserving instructor-authored material (\S\ref{sec:methods}). 
    \item In a between-subjects user study with 62 university students, we compared the effectiveness of AI-generated short-form reels against traditional long-form videos across learning outcomes, engagement, efficiency, and user trust (\S\ref{sec:results}). Results showed that learners using ReelsEd scored significantly higher on quizzes and completed tasks more quickly without any increase in cognitive load compared to long-form videos.
\end{enumerate}

We conclude by discussing how generative AI can be integrated into educational tools without sacrificing pedagogical clarity, and reflecting on design implications for AI-assisted microlearning to enhance learner agency and trust (\S\ref{sec:discussion}).
\section{Related Work}
\label{sec:related}

\begin{figure*}[t!]
    \centering
    \begin{subfigure}[b]{\textwidth}
        \centering
        \includegraphics[width=\textwidth]{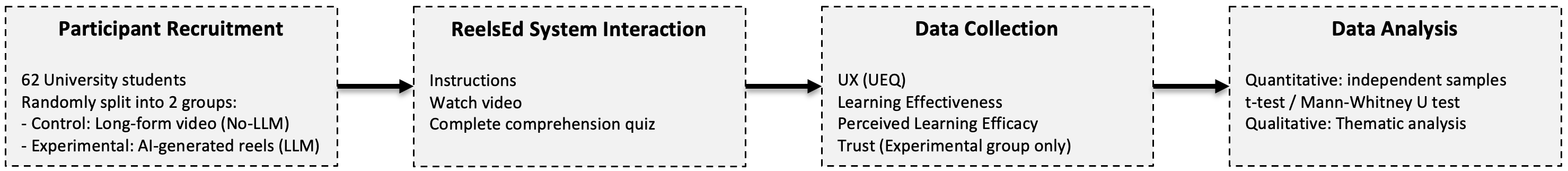}
    \end{subfigure}
    \caption{Overview of research methodology. Sixty-two university students were randomly assigned to either a control group (long-form video) or an experimental group (LLM-generated reels). Participants interacted with the ReelsEd system, watched their assigned video format, and completed a comprehension quiz. Data collection included UX (UEQ), learning effectiveness, perceived learning efficacy, and trust (for the experimental group only). Quantitative data were analyzed using t-tests or Mann-Whitney U tests, while qualitative responses were thematically analyzed.}
    \label{fig:methodology}
\end{figure*}

We surveyed various lines of research that our work draws upon, and grouped them into three main
areas: \emph{i)} microlearning; \emph{ii)} generative AI in education; and \emph{iii)} trust in AI-generated content.
\smallskip

\noindent\textbf{Microlearning in Education.} Microlearning (i.e., the use of bite-sized educational content) is used to boost engagement and support learning~\cite{kohnke2024microlearning, zhang2020designing}. TikTok, Instagram Reels, and YouTube Shorts are popular among Generation Z learners due to their brevity and visual appeal~\cite{conde2024microlearning, doloi2024reels, germanakos2019metacognitive}. Studies have shown that combining microlearning with social media can enhance learning outcomes by increasing satisfaction and improving engagement~\cite{denojean2024microlearning, jahnke2020unpacking}. For instance, TikTok-based university courses have been associated with strong student satisfaction and engagement~\cite{conde2024microlearning, escamilla2021incorporating}, with even educators becoming ``TikTok microcelebrities'' (i.e., those who use the platform for professional development~\cite{carpenter2024and}). Likewise, Instagram Reels influence content habits, with young adults favoring short and visually rich content~\cite{doloi2024reels}. Microlearning is often described as a flexible approach that enables learners to target relevant tasks efficiently~\cite{kohnke2024microlearning}. Consequently, researchers call for studying microlearning through platforms like TikTok and YouTube Shorts to expand access to educational content~\cite{aslan2024bite}. However, most existing work focuses on human-curated content, leaving limited empirical evidence on the pedagogical value and learner experience of AI-generated short-form videos.
\smallskip

\noindent\textbf{Generative AI in Education.} Generative AI has created new ways for generating and personalizing educational content~\cite{baidoo2023education, niu2023exercise, qadir2023engineering, mcdonald2025generative, liarokapis2024extended, liarokapis2024xr4ed, constantinides2024culturai}. Tools such as ChatGPT have prompted experiments with AI-driven teaching assistants, automated feedback, and content generation. Boumalek et al.~\cite{boumalek2024transforming} outlined how generative models can produce concise educational content through stepwise processes, boosting engagement and scalability. Reviews highlighted benefits such as improved accessibility, personalized support, and learning outcomes~\cite{bettayeb2024exploring, mittal2024review}. However, findings remain mixed. While some studies report comparable learning outcomes between AI- and human-created videos, learners often prefer human content for its relatability and engagement~\cite{netland2025comparing}. Concerns about factual accuracy, oversimplification, and narrative coherence persist~\cite{jensen2024generative, qadir2023engineering}. Ethical issues such as transparency and learner control are increasingly emphasized~\cite{cotton2024chatting, chan2023comprehensive, constantinides2024implications}, with scholars such as Brusilovsky advocating for AI systems that support human-AI collaboration through greater user agency~\cite{brusilovsky2024ai}. Despite these developments, little is known about how learners perceive and trust AI-generated short-form content. While LLMs can rapidly create microlearning materials, their impact on trust, perceived value, and overall learning experience remains underexplored.
\smallskip

\noindent\textbf{Trust in AI-Generated Content and Human–AI Interaction.} As AI tools become more integrated into education, trust in AI-generated content has emerged as a key concern~\cite{boumalek2024transforming, balakrishnan2021role, al2022understanding, kim2022perceived, fidas2023ensuring}. A central question is whether students trust AI-driven instruction as much as human-delivered support. Research on AI teaching assistants (AI TAs) suggests that trust is shaped by perceptions of ability, benevolence, and integrity~\cite{peng2024human, constantinides2023trustid}. These trust dimensions influence how learners engage with AI explanations, feedback, and summaries. Peng and Wan~\cite{peng2024human} found that trust in AI depends not only on performance quality, but also on communication style and perceived empathy, especially in complex or high-stakes tasks. Similarly, learners value transparency, authorship clarity, and a sense of control in AI-mediated learning environments~\cite{brusilovsky2024ai, portugal2023continuous}. Yet, these human-centered design aspects remain underexplored in educational AI systems and little is known whether learners trust AI-generated short-form video.
\smallskip
\section{Methods}
\label{sec:methods}

\begin{figure*}[t!]
    \centering
    \begin{subfigure}[b]{0.43\textwidth}
        \centering
        \includegraphics[width=\textwidth]{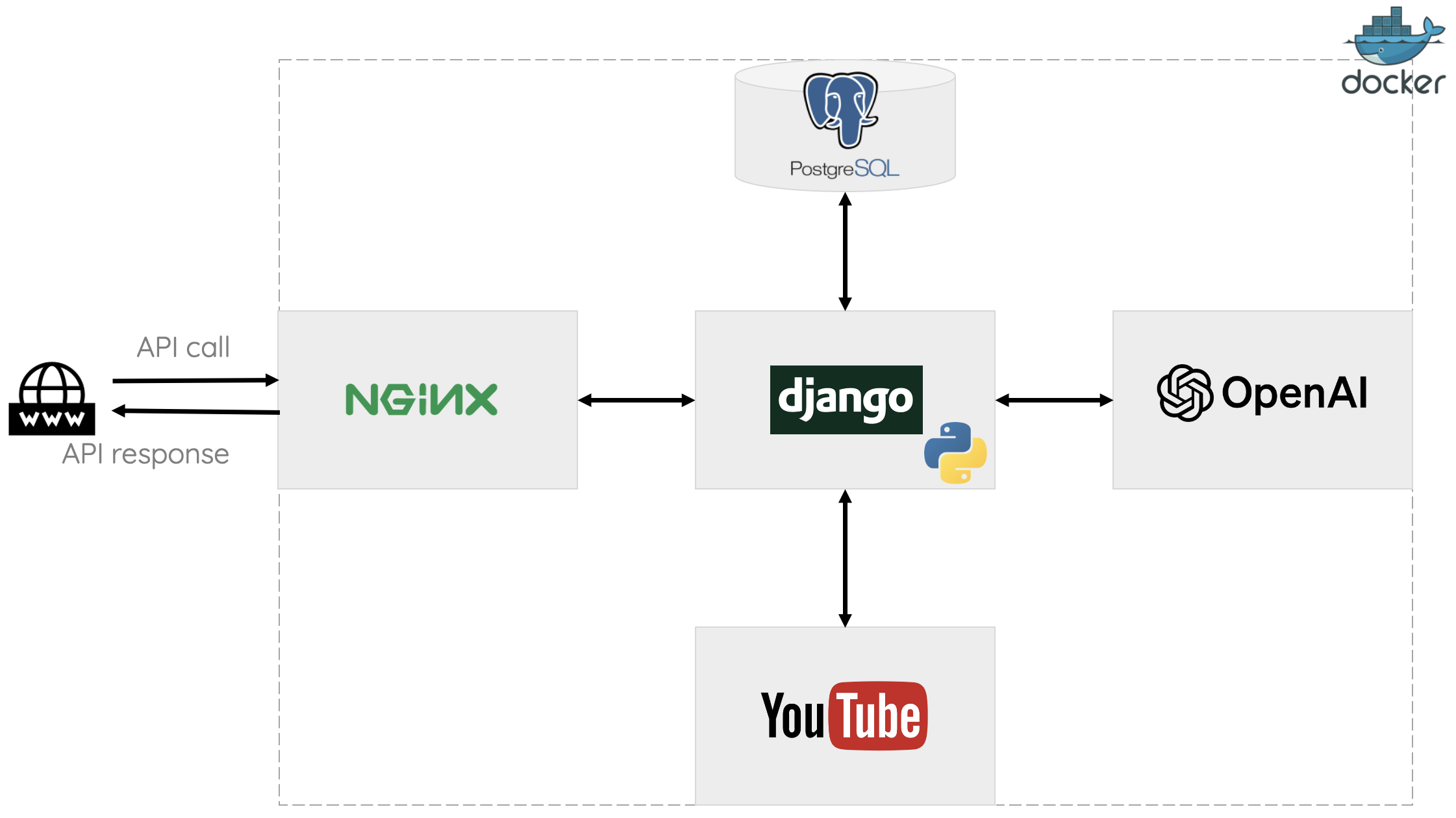}
        \caption{}
        \label{fig:ReelsEd_System_Architecture}
    \end{subfigure}
    \hfill
    \begin{subfigure}[b]{0.43\textwidth}
        \centering
        \includegraphics[width=\textwidth]{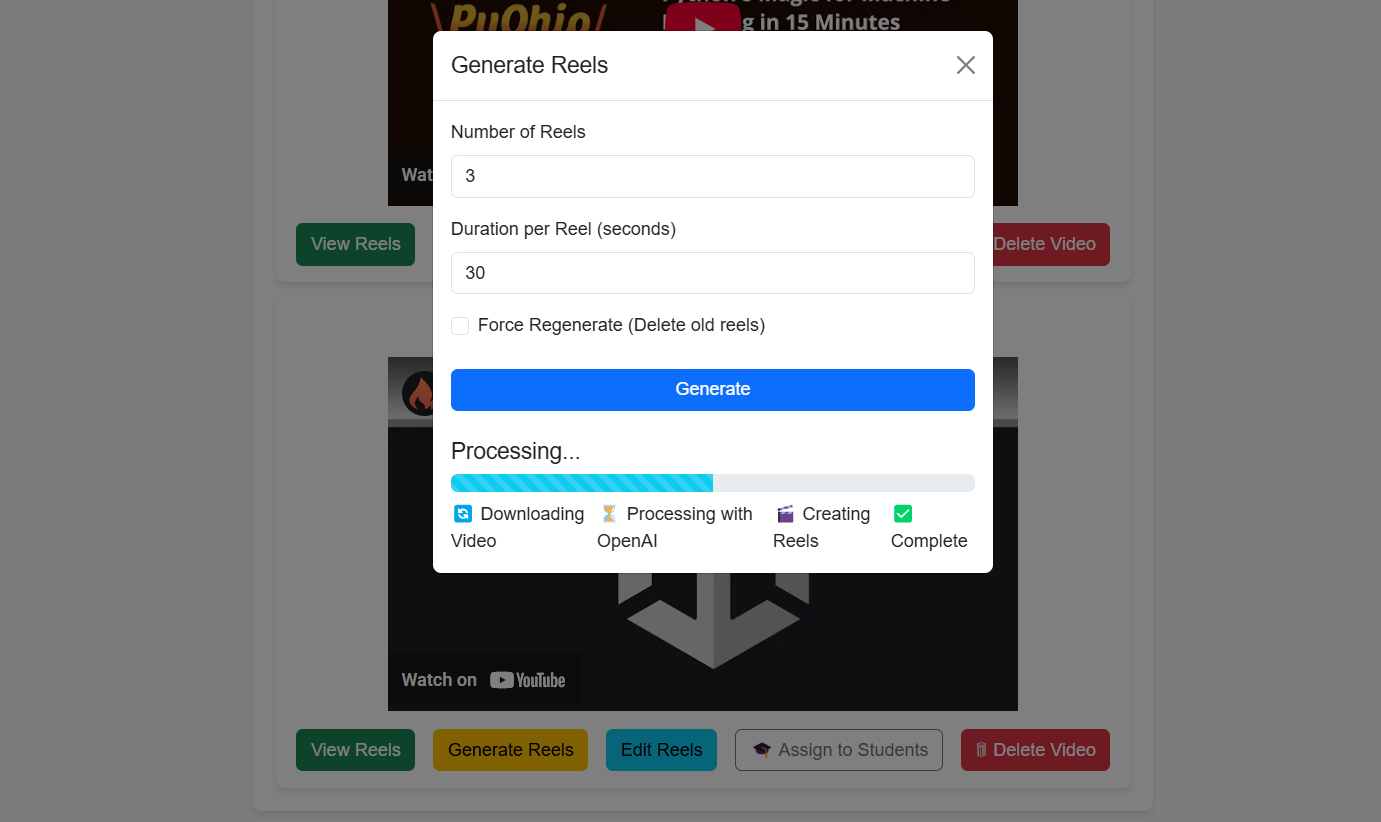}
        \caption{}
        \label{fig:Reels_Generation}
    \end{subfigure}
    \caption{Overview of the ReelsEd system and user interface. (a) The system architecture includes a Django-based web server that connects with PostgreSQL for data storage, OpenAI's GPT-4 API for content summarization, and YouTube for transcript extraction. NGINX handles API traffic, and the entire system is containerized using Docker. (b) Snapshot of the ReelsEd web interface where instructors can specify the number and duration of reels, triggering the automated generation pipeline. The UI reflects progress through stages such as downloading, LLM processing, and reel creation.}
    \label{fig:ReelsEd_Solution}
\end{figure*}

The study employed a between-subjects experimental design in which participants were randomly assigned into one of the two conditions: 1) long-form video (baseline), and 2) AI-Generated short-form video (ReelsEd). Each condition featured the same educational content, but formatted differently based on the experimental manipulation (Figure~\ref{fig:methodology}).

\subsection{Study Instruments}
\noindent\textbf{ReelsEd.} A web-based system, accompanied by a mobile application, that enables instructors to upload lecture videos and automatically generate AI-assisted short-form video reels using the OpenAI Application Programming Interface (API). The design of ReelsEd leverages principles derived from cognitive load theory~\cite{sweller1994cognitive} and microlearning pedagogy~\cite{taylor2022effects} to optimize learning efficiency and engagement by segmenting long videos into concise reels. Furthermore, to promote human-AI trust,~\cite{lee2004trust}, our design prioritizes pedagogical alignment with original instructor material.

A high-level overview of ReelsEd's architecture is illustrated in Figure \ref{fig:ReelsEd_System_Architecture}. It consists of a Django-based web application that communicates via API calls, providing seamless access to front-end clients. An NGINX server sits in front of the Django application to efficiently handle incoming requests, manage load balancing, and serve static assets. Data is securely stored in a PostgreSQL database for reliable and scalable management. The entire solution is containerized using Docker to ensure consistency across environments, and simplified deployment and maintenance.

The web-based system was built on Python 3.10.8 using the Django Framework, and the following Python libraries: \textit{i)} OpenAI for integrating GPT-4 (openai==1.63.0); \textit{ii)} YouTube downloader for downloading YouTube videos (yt-dlp==2025.01.26); \textit{iii)} YouTube transcript API for extracting video transcripts (youtube-transcript-api==1.0.3); and \textit{iv)} MoviePy for trimming of videos and generation of reels (moviepy==1.0.3). The mobile application was built using React Native (version 0.76.7) and Expo CLI (version 0.22.20). 

ReelsEd's automated reel generation follows a step-by-step process to ensure content coherence and pedagogical relevance. As shown in Figure \ref{fig:Reels_Generation}, it starts by extracting full transcripts from YouTube videos using the YouTube API. These transcripts are sent to GPT-4 for a multi-stage analysis: \textit{i) Key Moment Identification:} GPT-4 analyzes the transcript to find core concepts and key moments (as timestamps) based on a prompt that highlights the video’s learning goals. Instructors use the UI to set how many reels to create and how long each should be; \textit{ii) Segment Summarization and Labeling:} For each key moment, GPT-4 generates a short summary and a clear label. This step uses additional model calls to improve clarity and educational quality; and \textit{iii) Video Trimming and Assembly:} Based on the identified timestamps, MoviePy trims the original video into segments, which are then combined into a cohesive educational reel.

Instructors can review, edit, and assign the generated reels to students, who can then watch and rate them. Designed as an extensible client-server system, ReelsEd supports both instructors and students, enabling a streamlined workflow for AI-assisted video summarization and personalized learning experiences. The source code of ReelsEd is publicly available\footnote{ReelsEd Repository: \url{https://github.com/lstavrinu/ReelsEd-CHIGREECE-2025}}, to facilitate further research and reproducibility in the field of AI-assisted microlearning. The study utilized ReelsEd to generate short-form content for the experimental condition (\textit{i.e.}, the LLM group), demonstrating its potential for enhancing engagement and learning efficiency.
\smallskip

\noindent\textbf{Video Stimuli.} Four publicly available videos across different introductory topics were used: \textit{a)} \textit{Python}~\cite{LearnPythonVideo}; \textit{b)} \textit{C}~\cite{LearnCVideo}; \textit{c)} \textit{Java}~\cite{LearnJavaVideo}; and \textit{d)} \textit{Machine Learning}~\cite{LearnMLVideo}. These videos were selected based on three criteria: \textit{i)} comparable duration (10–15 minutes) to ensure consistency in cognitive load; \textit{ii)} structured educational delivery with verbal narration, which is essential for transcript-based summarization by LLMs~\cite{boumalek2024transforming}; and \textit{iii)} public availability and licensing that permitted academic use. These videos were used in two conditions: in the long-form condition, they were presented as a continuous 10–15-minute lecture video; in the AI-Generated (ReelsEd) condition, they were segmented into 5–6 AI-identified key moments (approximately 30–60 seconds each) using ReelsEd's summarization functionality.

To reduce bias and enhance validity, students were recruited from a diverse set of academic backgrounds, including disciplines in Computer and Data Sciences, Engineering, Natural Sciences, Mathematics, Social Sciences, and Business. To account for varying levels of prior experience, non-Computer Science students were assigned to one of the introductory programming topics (i.e., \textit{a)}, \textit{b)}, or \textit{c)}) to ensure accessibility, while Computer Science students were assigned to the more advanced machine learning topic \textit{d)} to match their prior knowledge and provide an appropriate level of challenge. The use of programming-related topics in such assessments has been employed in microlearning studies~\cite{polasek2019results, skalka2021conceptual} to ensure equity and enable meaningful comparisons of cognitive workload and engagement across diverse backgrounds.

\subsection{Measures}
\noindent\textbf{User Experience (UX)}: The short version of User Experience Questionnaire (UEQ) \cite{schrepp2014applying} was used to assess perceived usability, stimulation, and attractiveness across conditions. The UEQ captures the scales of \textit{pragmatic quality}, \textit{hedonic quality}, and \textit{overall quality}, through the dimensions of obstructive/supportive, complicated/easy, inefficient/efficient, confusing/clear, boring/exciting, not interesting/interesting, conventional/inventive, and usual/leading edge.

\noindent\textbf{Learning Effectiveness}: Participants completed a 6-item multiple-choice quiz designed to assess their recall and comprehension of the video content by consulting two university educators. Learning effectiveness was evaluated through three metrics: \textit{i)} the quiz score, measured as the percentage of correct answers; \textit{ii)} the quiz completion time, representing the total duration taken to complete the quiz; and \textit{iii)} the number of video revisits, which captured how often participants returned to the video content while answering quiz questions.

\noindent\textbf{Perceived Learning Efficacy}: To assess perceived learning efficacy, we used: \textit{i)} the Perceived Competence subscale from the Intrinsic Motivation Inventory, rated on a 7-point Likert scale (1 = Not at all true; 7 = Very true); \textit{ii)} the NASA Task Load Index, adapted to a 7-point Likert scale (1 = Very Low / Perfect; 7 = Very High / Failure); \textit{iii)} additional items evaluating the perceived learning effectiveness, including perceived understanding of the video topic, ability to retain key information, ability to maintain focus, and confidence in explaining the content to others; and \textit{iv)} items targeting the perceived learning experience of the short-form format itself, assessing whether the segmented presentation improved engagement, helped structure the material into manageable parts, supported information recall, and encouraged future use of similar formats. All items were rated on a 7-point Likert scale (1 = Not at all; 7 = Extremely).

\noindent\textbf{Trust in LLM-Generated Content}: For the ReelsEd condition, participants completed sixteen 7-point Likert scale items adapted from Jian et al.’s empirically derived trust in automation scale \cite{jian2000foundations}. These items assessed multiple dimensions of trust: perceptions of system integrity, reliability, security, and confidence, as well as skepticism, wariness, and concerns about deception or harmful intent. Additional custom items captured trust in the AI-generated reels focusing on their perceived accuracy, helpfulness for learning, and participants’ willingness to rely on them in the future. An open-ended question was also included to gather qualitative feedback on participants’ trust-related perceptions.

\noindent\textbf{Qualitative Data}: In both groups, participants were asked two open-ended questions to capture their reflections on the content delivery: Q1: ``What did you find most helpful in how the content was delivered?''; and, Q2: ``Was anything missing that would have helped you learn better?'' These questions captured learners' experiences and preferences, and allowed us to triangulate the quantitative results.

\subsection{Sampling and Procedure}
A total of 62 university undergraduate and postgraduate students from an Eastern European country participated in the study ($M_{age}$ = 22.09, $SD_{age}$ = 1.37; 27 female, 35 male, 0 non-binary). Students were recruited through email invitations and in-class announcements made by colleagues.
Participants had no relationship with researchers to avoid bias. Participation was voluntary and participants could opt-out from the study at any time. We adopted the university’s human research protocol for conducting user studies, which considers users’ privacy, confidentiality, and anonymity of users. Accordingly, participants were asked to sign a consent form prior to their participation in the study. Participants signed up for a scheduled in-lab session and initially completed a short pre-study questionnaire including demographics and familiarity with short-form learning video content. They were then randomly assigned to one of the two groups (LLM vs. No-LLM). Participants watched their assigned video(s), completed the comprehension quiz, and filled out the post-study questionnaires assessing UX and perceived learning efficacy. In the case of the LLM-group, participants also completed a questionnaire regarding trust in AI-generated content. During the participants' interactions, the researcher captured the time needed to complete the comprehension quiz, as well as notes on participants' revisits on the video(s). The total duration of participation was approximately 20–25 minutes.

\subsection{Data Analysis}
To compare the LLM and No-LLM groups across different outcome measures, we first assessed the normality of data distributions using the Shapiro-Wilk test. If both groups met the assumption of normality (p > 0.05), we proceeded with an independent samples t-test. Otherwise, we employed the Mann-Whitney U test, a non-parametric alternative that does not require normally distributed data. For each variable, we report the mean and standard deviation for both groups, along with the test statistic and p-value. A significance threshold of p < 0.05 determined statistical significance. Also, qualitative responses were thematically analyzed to identify recurring themes related to trust, engagement, and suggestions for improvement. Two researchers conducted inductive thematic analysis~\cite{braun2006using} on the open-ended responses, grouped initial codes into broader themes and resolved discrepancies through discussion to ensure reliability.
\section{Results}
\label{sec:results}
\subsection{User Experience Questionnaire} The overall quality of the control group (\textit{i.e.}, No-LLM) was 1.810 (pragmatic quality: 2.024; hedonic quality: 1.597), while for the experimental group (\textit{i.e.}, LLM) it was 1.923 (pragmatic quality: 2.339; hedonic quality: 1.508).

\subsection{Learning Effectiveness}
Learning effectiveness was evaluated through quiz scores, quiz completion time, and number of video revisits.
\smallskip

\begin{figure}[t!]
    \centering
    \includegraphics[width=\columnwidth]{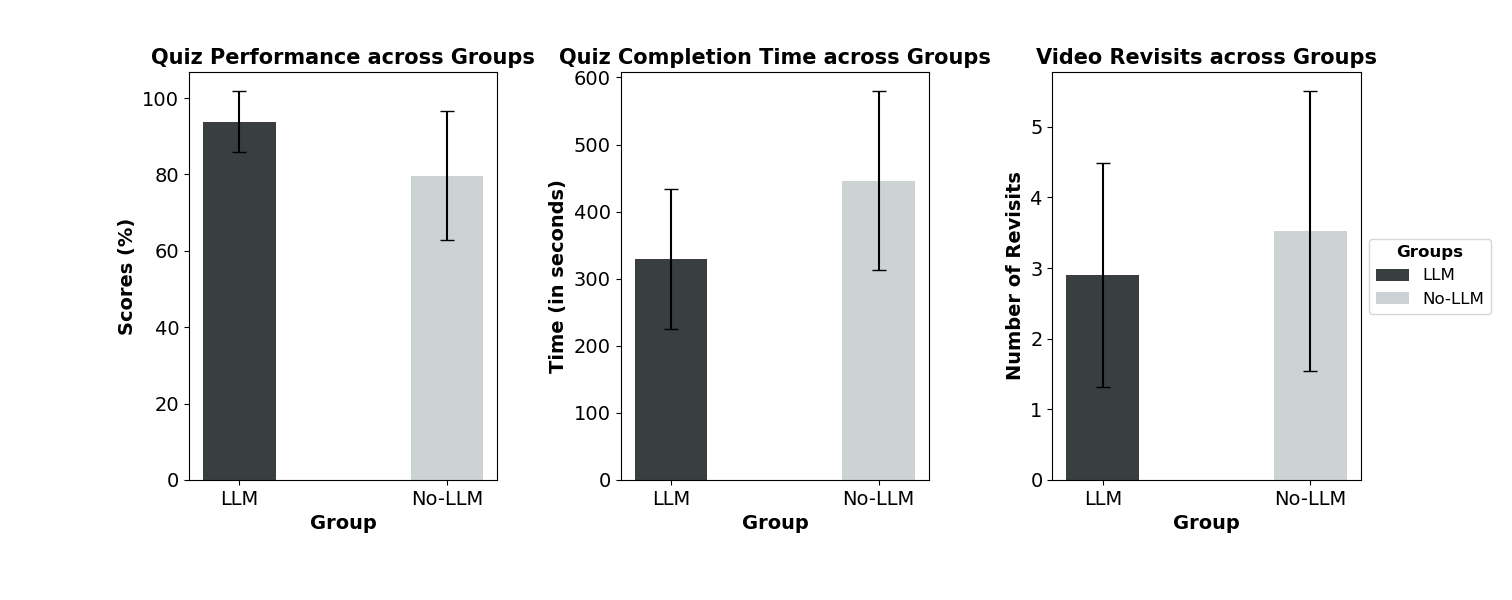}
    \caption{Comparison of learning effectiveness metrics between the LLM-generated short-form video group and the traditional long-form video group. The LLM group achieved higher quiz scores (left), completed the quiz in less time (middle), and had fewer video revisits (right), although the difference in revisits was not statistically significant. Error bars represent standard deviation.}
    \label{fig:learning_effectiveness_results}
\end{figure}

\noindent{\textbf{Quiz Scores:} Students in the LLM group scored significantly higher than those in the No-LLM group. The mean quiz score for the LLM group was M = 93.85\%, SD = 7.89, compared to M = 79.72\%, SD = 16.98 for the No-LLM group, as illustrated in Figure \ref{fig:learning_effectiveness_results} (left). A Mann-Whitney U test indicated that this difference was statistically significant, U = 736.50, p = 0.0001, suggesting that the LLM-assisted microlearning format enhanced quiz performance.}
\smallskip

\noindent{\textbf{Quiz Completion Time:} Students in the LLM group completed the quiz significantly faster than those in the No-LLM group. The mean quiz duration for the LLM group was M = 328.77 seconds (5:28), SD = 104.26, compared to M = 446.23 seconds (7:26), SD = 132.87 for the No-LLM group, as illustrated in Figure \ref{fig:learning_effectiveness_results} (middle). A Mann-Whitney U test indicated that this difference was statistically significant, U = 219.00, p = 0.0002, suggesting that the LLM-assisted microlearning format improved the efficiency of quiz completion.}
\smallskip

\noindent{\textbf{Number of Video Revisits:} Students in the LLM group revisited the video fewer times than those in the No-LLM group, but this difference was not statistically significant. The mean number of video revisits for the LLM group was M = 2.90, SD = 1.58, compared to M = 3.52, SD = 1.98 for the No-LLM group, as illustrated in Figure \ref{fig:learning_effectiveness_results} (right). A Mann-Whitney U test indicated that this difference was not statistically significant, U = 402.50, p = 0.2679, suggesting that the LLM-assisted microlearning format did not significantly affect the number of video revisits.}

The results indicate that the LLM-assisted microlearning format significantly improved both quiz performance and efficiency. Students in the LLM group scored significantly higher on the quiz and completed it in less time compared to those in the No-LLM group, highlighting the format’s effectiveness in supporting comprehension and recall. Although the LLM group revisited the videos slightly less than the No-LLM group, the difference was not statistically significant. These improvements in performance were likely driven by the format’s instructional quality rather than by repeated viewing. In other words, it's not about watching more, but learning better, as the microlearning approach helped learners grasp and retain key concepts more effectively on the first try.

\subsection{Perceived Learning Efficacy}
Our results indicate that AI-generated short-form videos positively impact perceived competence, learning effectiveness, and engagement without increasing cognitive load. We found significant differences between the No-LLM and LLM groups across several dimensions (Table ~\ref{tableresultssummary}). In terms of perceived competence, participants in the LLM group consistently rated themselves as statistically more competent, skilled, and satisfied with their performance compared to the No-LLM group. For the NASA Task Load Index, we found no significant differences in mental demand, effort, or frustration between the two formats. Nonetheless, the LLM group reported slightly lower mental effort and emotional strain compared to the No-LLM group. For the item \textit{``How successful were you in accomplishing what you were asked to do?''}, the LLM group reported a slightly lower average score than the No-LLM group, though the difference was not statistically significant. This suggests that some students may have felt less successful because the AI-generated reels were concise and easy to follow. Even with high quiz scores, the task may have felt less demanding, which led to lower perceived success. Regarding perceived learning effectiveness, participants found the short-form format more helpful for retaining key information, maintaining focus, and improving their ability to explain the topic to others, with all related questions showing strong statistical significance. Finally, under perceived learning experience, participants in the experimental group reported higher engagement, improved content breakdown, and a greater preference for future learning in this format, highlighting the advantages of modular, accessible content for enhanced learner experience.

\begin{table*}[t!]
    \centering
      \caption{Perceived learning efficacy measures across two groups (No-LLM \emph{vs.} LLM) and four dimensions: \textit{perceived competence}, \textit{task load}, \textit{perceived learning effectiveness}, and \textit{perceived learning experience}. The \textit{Dimension} column reflects the conceptual categories assessed through questionnaire items. Values represent $\mu$ and $\sigma$, with statistically significant results in bold.}
    \label{tableresultssummary}
    
    \resizebox{\textwidth}{!}{%
        \begin{tabular}{c|c|p{2.5cm}|p{2.5cm}}
            \hline
            \multicolumn{1}{c|}{\textbf{Dimension}} & \multicolumn{1}{c|}{\textbf{Question}} & \multicolumn{1}{c|}{\shortstack{\textbf{No-LLM vs. LLM Group}\\($\mu$ $\pm$ $\sigma$)}} & \multicolumn{1}{c}{\shortstack{\textbf{Test Statistic}}} \\
            \hline
            \multirow{6}{*}{\textit{Perceived Competence}} & \multicolumn{1}{c|}{\textbf{I think I am pretty good at this activity.}} & \multicolumn{1}{c|}{5.32 $\pm$ 1.35 vs. \textbf{6.03} $\pm$ \textbf{0.84}}  & \multicolumn{1}{c}{U = 624.50, \textbf{p = 0.0357}} \\
            \cline{2-4}
             & \multicolumn{1}{c|}{I think I did pretty well at this activity, compared to other students.} & \multicolumn{1}{c|}{5.16 $\pm$ 1.49 vs. 5.87 $\pm$ 0.99}  & \multicolumn{1}{c}{U = 610.00, p = 0.0615} \\
            \cline{2-4}
             & \multicolumn{1}{c|}{\textbf{After working at this activity for a while, I felt pretty competent.}} & \multicolumn{1}{c|}{5.29 $\pm$ 1.40 vs. \textbf{5.97} $\pm$ \textbf{1.11}}  & \multicolumn{1}{c}{U = 626.50, \textbf{p = 0.0336}} \\
             \cline{2-4}
             & \multicolumn{1}{c|}{\textbf{I am satisfied with my performance at this task.}} & \multicolumn{1}{c|}{5.42 $\pm$ 1.46 vs. \textbf{6.52} $\pm$ \textbf{0.68}}  & \multicolumn{1}{c}{U = 714.00, \textbf{p = 0.0005}} \\
             \cline{2-4}
             & \multicolumn{1}{c|}{\textbf{I was pretty skilled at this activity.}} & \multicolumn{1}{c|}{5.03 $\pm$ 1.54 vs. \textbf{6.10} $\pm$ \textbf{0.87}}  & \multicolumn{1}{c}{U = 690.00, \textbf{p = 0.0023}} \\
             \cline{2-4}
             & \multicolumn{1}{c|}{This was an activity that I couldn't do very well.} & \multicolumn{1}{c|}{3.87 $\pm$ 1.73 vs. 3.71 $\pm$ 2.13}  & \multicolumn{1}{c}{U = 453.00, p = 0.7007} \\
            \hline
            \multirow{6}{*}{\textit{Task Load Index}} & \multicolumn{1}{c|}{How mentally demanding was the task?} & \multicolumn{1}{c|}{4.10 $\pm$ 1.51 vs. 3.71 $\pm$ 1.75}  & \multicolumn{1}{c}{U = 426.50, p=0.4426} \\
            \cline{2-4}
             & \multicolumn{1}{c|}{How physically demanding was the task?} & \multicolumn{1}{c|}{1.68 $\pm$ 1.11 vs. 1.58 $\pm$ 0.89}  & \multicolumn{1}{c}{U = 488.50, p=0.9013} \\
            \cline{2-4}
             & \multicolumn{1}{c|}{How hurried or rushed was the pace of the task?} & \multicolumn{1}{c|}{3.52 $\pm$ 1.39 vs. 3.26 $\pm$ 1.34}  & \multicolumn{1}{c}{U = 449.00, p=0.6467} \\
             \cline{2-4}
             & \multicolumn{1}{c|}{How successful were you in accomplishing what you were asked to do?} & \multicolumn{1}{c|}{3.39 $\pm$ 1.91 vs. 2.81 $\pm$ 2.33}  & \multicolumn{1}{c}{U = 376.00, p=0.1325} \\
             \cline{2-4}
             & \multicolumn{1}{c|}{How hard did you have to work to accomplish your level of performance?} & \multicolumn{1}{c|}{4.29 $\pm$ 1.66 vs. 3.58 $\pm$ 1.78}  & \multicolumn{1}{c}{U = 366.00, p=0.1028} \\
             \cline{2-4}
             & \multicolumn{1}{c|}{How insecure, discouraged, irritated, stressed, and annoyed were you?} & \multicolumn{1}{c|}{2.81 $\pm$ 1.74 vs. 1.97 $\pm$ 1.28}  & \multicolumn{1}{c}{U = 356.00, p=0.0677} \\
            \hline
            \multirow{4}{*}{\textit{Perceived Learning Effectiveness}} & \multicolumn{1}{c|}{\textbf{I was able to understand the topic well through these videos.}} & \multicolumn{1}{c|}{4.71 $\pm$ 1.46 vs. \textbf{6.29} $\pm$ \textbf{0.92}}  & \multicolumn{1}{c}{U = 857.50, \textbf{p<0.0001}} \\
            \cline{2-4}
             & \multicolumn{1}{c|}{\textbf{The short-form video format helped me retain key information.}} & \multicolumn{1}{c|}{4.81 $\pm$ 1.55 vs. \textbf{6.29} $\pm$ \textbf{1.22}}  & \multicolumn{1}{c}{U = 805.00, \textbf{p<0.0001}} \\
            \cline{2-4}
             & \multicolumn{1}{c|}{\textbf{This format helped me focus better than traditional video lectures.}} & \multicolumn{1}{c|}{4.97 $\pm$ 1.79 vs. \textbf{6.03} $\pm$ \textbf{1.38}}  & \multicolumn{1}{c}{U = 697.00, \textbf{p=0.0099}} \\
             \cline{2-4}
             & \multicolumn{1}{c|}{\textbf{I feel more confident explaining this topic to others now.}} & \multicolumn{1}{c|}{4.74 $\pm$ 1.80 vs. \textbf{5.65} $\pm$ \textbf{1.18}}  & \multicolumn{1}{c}{U = 661.00, \textbf{p=0.0419}} \\
            \hline
            \multirow{5}{*}{\textit{Perceived Learning Experience}} & \multicolumn{1}{c|}{\textbf{The videos helped me remember key points better than a full lecture.}} & \multicolumn{1}{c|}{4.81 $\pm$ 1.49 vs. \textbf{6.19} $\pm$ \textbf{1.11}}  & \multicolumn{1}{c}{U = 732.50, \textbf{p=0.0002}} \\
            \cline{2-4}
             & \multicolumn{1}{c|}{\textbf{The format helped break down the topic into manageable parts. }} & \multicolumn{1}{c|}{4.45 $\pm$ 1.65 vs. \textbf{6.45} $\pm$ \textbf{0.81}}  & \multicolumn{1}{c}{U = 818.00, \textbf{p<0.0001}} \\
            \cline{2-4}
             & \multicolumn{1}{c|}{\textbf{I would prefer to learn future topics using this format.}} & \multicolumn{1}{c|}{4.26 $\pm$ 2.02 vs. \textbf{5.87} $\pm$ \textbf{1.12}}  & \multicolumn{1}{c}{U = 706.50, \textbf{p=0.0011}} \\
            \cline{2-4}            
             & \multicolumn{1}{c|}{\textbf{This format made it easier to revisit important concepts.}} & \multicolumn{1}{c|}{4.00 $\pm$ 2.03 vs. \textbf{6.55} $\pm$ \textbf{0.77}}  & \multicolumn{1}{c}{U = 843.50, \textbf{p<0.0001}} \\
             \cline{2-4}
             & \multicolumn{1}{c|}{\textbf{I felt more engaged with this format compared to traditional lectures.}} & \multicolumn{1}{c|}{4.13 $\pm$ 2.08 vs. \textbf{6.06} $\pm$ \textbf{1.46}}  & \multicolumn{1}{c}{U = 739.00, \textbf{p=0.0002}} \\
            \hline
        \end{tabular}
    }
\end{table*}

\subsection{Qualitative Analysis}

\subsubsection{\textbf{Learning Experience}}
To triangulate the quantitative findings, we analyzed the two open-ended questions.

\noindent\textbf{Q1: ``What did you find most helpful in how the content was delivered?''} Analysis of participants' responses revealed that the LLM-generated short-form videos were preferred for their efficiency, ease of navigation, and structured content delivery. Learners in this group found that the bite-sized format made information easier to process, retain, and revisit. The ability to navigate reels quickly and avoid unnecessary details was highly appreciated, making the format more engaging and user-friendly. Indicative responses include: \textit{"Each reel did only a specific task and i could understand it better before watching the next one." \(\sim \) P1}; \textit{"Delivered and explained the key points to help me understand." \(\sim \) P2}; \textit{"Reels' content is placed in order, based on the steps of training and testing a neural network." \(\sim \) P5}; \textit{"That it was easy to navigate and change the reel." \(\sim \) P6}; \textit{"The short videos format made the information more memorable and easier to understand." \(\sim \) P8}; \textit{"The fact that everything was divided into topics made learning easier and clearer." \(\sim \) P13}; \textit{"I knew exactly were to find what compared watching it in a full video (from previous experience)" \(\sim \) P15}. Conversely, the No-LLM group found value in coherence and logical sectioning, but navigation issues and engagement challenges were key concerns. While the ability to rewatch content was useful, some learners struggled with finding information easily and felt the video lacked an engaging introduction. Indicative responses include: \textit{"The way its concepts were separated and clear." \(\sim \) P2}; \textit{"Not helpful, [...] the video should explain better what it is all about and not jump into the process directly." \(\sim \) P4}; \textit{"The video is too long and I couldn't find what I was searching for very easy" \(\sim \) P11}.

\noindent\textbf{Q2: ``Was anything missing that would have
helped you learn better?''} While participants in the LLM group generally appreciated the short-form format and found it effective, they identified areas for improvement mostly related to navigational features, content scaffolding, and enhanced note-taking tools. Features like a content gallery, subtitles, and note-friendly resources (scripts or key point summaries) could  enhance the user experience. Their feedback reflects a desire to optimize the delivery and usability, not the core content, which they found helpful. Indicative responses include: \textit{"A gallery with a title or a thumbnail for each small clip in a general view so I can see all the clips at once [...]." \(\sim \) P11}; \textit{"Grid of videos." \(\sim \) P14}; \textit{"Gallery of the reels so I can choose easily" \(\sim \) P17}.

The long-form format appeared to be more overwhelming and less efficient. Feedback suggests learners were looking for shorter, clearer, and more modular content that would help them focus and reduce cognitive load. Navigation within the video and lack of actionable practice were also issues. Overall, their responses underline the advantages of a microlearning approach, especially for comprehension, retention, and a sense of control over their own learning process. Indicative responses include: \textit{"Shorter format." \(\sim \) P4}; \textit{"This format should have been shorter." \(\sim \) P7}; \textit{"Smaller video." \(\sim \) P8}; \textit{"It has a lot of new information I couldn't handle." \(\sim \) P13}; \textit{"I could not find easy the part I wanted for the quiz." \(\sim \) P14}.

\subsubsection{\textbf{Trust in LLM-Generated Content}} To assess participants' trust in the LLM-generated content, we analyzed a series of items exclusively to the LLM group. The results, reported as Mean $\pm$ Standard Deviation, indicate generally high levels of trust and perceived reliability in the system. Participants reported strong agreement with positively framed trust statements, such as \textit{"I can trust the system"} (6.03 $\pm$ 0.91), \textit{"The system is dependable"} (5.90 $\pm$ 1.30), \textit{"The system is reliable"} (5.81 $\pm$ 1.08), and \textit{"The system has integrity"} (5.68 $\pm$ 1.14). Additionally, participants found the AI-generated reels to be trustworthy (6.29 $\pm$ 0.78) and accurate (6.32 $\pm$ 0.79), and expressed confidence in using the system for future learning (5.90 $\pm$ 0.98). Conversely, responses to negatively framed statements reflected low levels of distrust. Items such as \textit{"The system is deceptive"} (2.29 $\pm$ 1.46), \textit{"The system behaves in an underhanded manner"} (2.13 $\pm$ 1.33), \textit{"I am suspicious of the system's intent, action, or outputs"} (2.16 $\pm$ 1.77), and \textit{"The system’s actions will have a harmful or injurious outcome"} (1.84 $\pm$ 1.42) received low agreement, indicating limited concern regarding deception or harmful intent.

Despite the overall high trust, there remained a moderate level of skepticism (3.13 $\pm$ 2.19) and familiarity with the system (5.35 $\pm$ 1.40), suggesting that while participants trusted the AI-generated content, there may still be room to improve system transparency. These findings suggest that participants perceived the LLM-generated microlearning content as credible and suitable for educational purposes, with minimal apprehension regarding its trustworthiness.
\section{Discussion}
\label{sec:discussion}
In this study, we provided empirical evidence that LLM-generated microlearning content can improve learning effectiveness, efficiency, and learner satisfaction. Our findings highlight the potential of generative AI to support human learning by offering more accessible and manageable formats for instructional content, and discuss its implications. Next, we discuss three main implications about the improved learning performance and efficiency, the learner experience and perceived effectiveness, and trust in LLM-generated content. We conclude with the limitations and future directions.
\smallskip

\noindent\textbf{Implications.} 
Participants in the LLM group achieved significantly higher quiz scores and completed the quiz in less time compared to those in the No-LLM group. These results suggest that short-form, AI-generated videos not only improved comprehension but also enabled learners to apply their understanding more efficiently. Importantly, the number of video revisits was not significantly different between groups, indicating that the performance benefits were not due to repeated exposure, but rather the clarity and focus of the content itself. These outcomes reflect core advantages of the microlearning format, particularly its alignment with cognitive load theory and principles of chunked learning. Our findings reinforce claims from prior work that short-form content enhances engagement and lowers cognitive load~\cite{conde2024microlearning, denojean2024microlearning, kohnke2024microlearning}. By breaking down complex topics into brief, coherent segments, learners could grasp and retain information more effectively without becoming overwhelmed.

Participants in the LLM group reported greater perceived competence, higher engagement, and more positive evaluations of the content’s usefulness for retaining and explaining key concepts. This suggests that they not only learned effectively but also felt confident in their understanding. In their qualitative responses, learners highlighted the structured delivery, ease of navigation, and focused nature of the short-form reels. The ability to quickly access specific information and avoid unnecessary detail was widely appreciated. In contrast, the No-LLM group noted difficulties with finding content in the long-form video and expressed a desire for a more modular presentation. These insights indicate that learners value content formats that offer a greater sense of control and better alignment with their learning preferences~\cite{boumalek2024transforming, bettayeb2024exploring, mittal2024review}, features that LLM-generated microlearning appears well-suited to provide.

Trust is a central factor in the adoption of AI systems in educational contexts. Results from the LLM group showed high levels of trust in the system’s reliability, accuracy, and integrity. Learners generally viewed the AI-generated content as dependable and suitable for learning, with minimal concerns around deception or harmful outcomes. While trust levels were strong overall, responses also indicated moderate levels of skepticism and a lack of familiarity with the system. Although learners accepted the AI-generated content, future iterations could benefit from more transparency in how content is created and features that help users understand the model’s role and limitations~\cite{brusilovsky2024ai, boumalek2024transforming, penalvo2024safe}.
\smallskip

\noindent\textbf{Limitations and Future Work.} 
Our study has limitations that call for future research efforts. The participant pool, though balanced in terms of group assignment, was limited in size and geographic diversity, consisting of 62 university students from an Eastern European country. Also, the video content was limited to introductory programming topics from YouTube, primarily due to its available transcript API. However, the system’s flexible architecture can support non-YouTube content through speech-to-text integration. Future work should include larger and more diverse learner populations to examine how cultural, linguistic, or disciplinary differences (e.g., humanities or arts where content nuance is complex) shape experiences with AI-generated educational content.

While our findings showed no increase in cognitive load, the long-term effects of using short-form formats on knowledge retention or the persistence of engagement benefits remain unclear and warrant longitudinal investigation. Furthermore, the observed positive experience may be partly due to the novelty of AI-generated content, which future work should test for long-term persistence as learners become more familiar with it. Also, using short quizzes helped reduce participant burden, but limited how well we could measure overall learning. Future studies could use more comprehensive and varied assessments. Also, while microlearning has benefits, AI-generated summaries can oversimplify, miss key context, or include errors. Future work should enable instructors review and improve these summaries to ensure deeper understanding.

Qualitative feedback also revealed areas where the ReelsEd interface could be improved (e.g., by adding subtitles, a gallery view for navigation, and support for note-taking). Future qualitative studies should explore negative perceptions to provide a balanced view of the user experience. For instance, some participants in the LLM group felt less successful despite high scores, possibly due to the reels’ efficient and concise format. This gap between perceived effort and actual learning should be examined further. Nonetheless, the qualitative insights and participants' recommendations point to opportunities for future design iterations. Such enhancements could make AI-assisted microlearning more accessible and adaptable to different learner needs. More broadly, we see potential in exploring how user trust evolves over time with repeated exposure to LLM-generated content, and how such systems might better support learner agency through personalization, transparency, and feedback mechanisms embedded in the learning experience.

\section{Conclusion}
\label{sec:conclusion}
In this study, we contribute to the growing body of research at the intersection of HCI, education, and generative AI by demonstrating that LLM-generated short-form content can be both pedagogically effective and well-received by learners. Rather than viewing generative AI as a replacement for traditional instruction, our findings suggest it can augment educational experiences by offering accessible, modular, and trustworthy learning formats. As microlearning becomes increasingly embedded in digital education, we argue that the design of AI-assisted tools must balance automation with learner control, transparency, and contextual relevance in a way that pushes forward a human-centered agenda in AI-enhanced learning environments. However, our findings are based on a specific context and should not be extrapolated across domains without further validation.

\begin{acks}
The work has received funding from the European Union’s Horizon 2020 Research and Innovation Programme Grant Agreement No. 739578 and the Government of the Republic of Cyprus through the Deputy Ministry of Research, Innovation and Digital Policy. It has also received funding from the European Union under grant agreement No. 101093159. Views and opinions expressed are those of the authors and do not necessarily reflect those of the European Union. Neither the European Union nor the granting authority can be held responsible for them.
\end{acks}

\newpage

\bibliographystyle{ACM-Reference-Format}
\bibliography{main}

\end{document}